\documentclass[apj,a4paper,12pt,useAMS]{emulateapj} 

\usepackage{graphicx,graphics}

\usepackage{natbib}

\newcommand{\simgt}{\lower.5ex\hbox{$\; \buildrel > \over \sim \;$}}
\newcommand{\simlt}{\lower.5ex\hbox{$\; \buildrel < \over \sim \;$}}

\def\singlebond{\@makechembond\@ne}
\def\doublebond{\@makechembond\tw@}
\def\triplebond{\@makechembond\thr@@}

\shortauthors{Liao et al.}

\shorttitle{AMiBA SZE Properties and Scaling Relations}

\begin{document}

\title{
AMiBA: Sunyaev-Zel'dovich Effect Derived Properties\\ 
and Scaling Relations of Massive Galaxy Clusters
}
\author{
Yu-Wei~Liao\altaffilmark{1,2}, Jiun-Huei~Proty~Wu\altaffilmark{1},
Paul~T.~P.~Ho\altaffilmark{2,3}, Chih-Wei~Locutus~Huang\altaffilmark{1},
Patrick~M.~Koch\altaffilmark{2}, 
Kai-Yang~Lin\altaffilmark{2,1}, Guo-Chin~Liu\altaffilmark{2,4}, 
Sandor~M.~Molnar\altaffilmark{2}, Hiroaki~Nishioka\altaffilmark{2},
Keiichi~Umetsu\altaffilmark{2}, Fu-Cheng~Wang\altaffilmark{1},
Pablo~Altamirano\altaffilmark{2}, Mark~Birkinshaw\altaffilmark{5},
Chia-Hao~Chang\altaffilmark{2}, Shu-Hao~Chang\altaffilmark{2},
Su-Wei~Chang\altaffilmark{2}, Ming-Tang~Chen\altaffilmark{2},
Tzihong~Chiueh\altaffilmark{1}, Chih-Chiang~Han\altaffilmark{2},
Yau-De~Huang\altaffilmark{2}, Yuh-Jing~Hwang\altaffilmark{2}, 
Homin~Jiang\altaffilmark{2}, Michael~Kesteven\altaffilmark{6},
Derek~Y.~Kubo\altaffilmark{2},
Chao-Te~Li\altaffilmark{2}, Pierre~Martin-Cocher\altaffilmark{2},
Peter~Oshiro\altaffilmark{2}, Philippe~Raffin\altaffilmark{2},
Tashun~Wei\altaffilmark{2}, Warwick~Wilson\altaffilmark{6}}

\altaffiltext{1}{Department of Physics, Institute of Astrophysics, \& Center
for Theoretical Sciences, National Taiwan University, Taipei 10617, Taiwan}
\altaffiltext{2}{Institute of Astronomy and Astrophysics, Academia Sinica,
P.~O.~Box 23-141, Taipei 10617, Taiwan}
\altaffiltext{3}{Harvard-Smithsonian Center for Astrophysics, 60 Garden
Street, Cambridge, MA 02138, USA}
\altaffiltext{4}{Department of Physics, Tamkang University, 251-37 Tamsui,
Taipei County, Taiwan}
\altaffiltext{5}{University of Bristol, Tyndall Avenue, Bristol BS8 1TL, UK}
\altaffiltext{6}{Australia Telescope National Facility, P.O.Box 76, Epping NSW 1710, Australia}

\begin{abstract}
The Sunyaev-Zel'dovich Effect (SZE) has been observed toward six massive
galaxy clusters, at redshifts $0.091{\leq}z{\leq}0.322$ in the 86-102 GHz band with the Y. T. Lee Array for Microwave 
Background Anisotropy (AMiBA). We modify an iterative method, based on the isothermal $\beta$-models, to derive the electron temperature $T_{\rm e}$, total mass $M_{\rm t}$,  
gas mass $M_{\rm g}$, and integrated Compton $Y$ within $r_{2500}$, from the AMiBA SZE data.
Non-isothermal universal temperature profile (UTP) $\beta$ models are also considered in this paper. 
These results are in good agreement with those deduced from other observations.
We also investigate the embedded scaling relations, due to the assumptions that have been made
in the method we adopted, between these purely SZE-deduced $T_{\rm e}$, $M_{\rm t}$, $M_{\rm g}$ and $Y$.
Our results suggest that 
cluster properties may be measurable with SZE observations alone.
However, the assumptions built into the pure-SZE method bias the results of 
scaling relation estimations and need further study.
\end{abstract}


\section{Introduction}\label{sec:intro}
The Sunyaev-Zel'dovich Effect (SZE) is an useful tool for studies of galaxy clusters.
This distortion of the Cosmic Microwave Background (CMB) is caused by
the inverse Compton scattering by high energy electrons as the CMB propagates
through the hot plasma of galaxy clusters \citep{SZ1972}. The SZE signal
is essentially redshift independent, making it particularly useful for
determining the evolution of large-scale structure.  

For upcoming SZE cluster surveys \citep{Ruhl2004,Fowler2004,Kaneko2006,Ho2008}, it is important to investigate
the relations between SZE flux density and other cluster properties, such as mass, temperature, and gas fraction.
By assuming that the evolution of clusters is dominated by self-similar gravitational processes,
we can predict simple power law relations between integrated Compton $Y$ and other cluster properties \citep{Kasier1986}.
Strong correlations between integrated SZE flux and the mass of clusters are also suggested by
numerical simulations \citep{dasilva2004,Motl2005,Nagai2006}. 
These relations imply
the possibility of determining the masses and temperatures of clusters, and investigating cluster evolution at high redshift, 
with SZE observation data alone.

\citet{Joy2001} and \citet{Bonamente2007} demonstrated an iterative approach based on the isothermal $\beta$ model to estimate the values of electron temperature $T_e$, total mass $M_t$, gas mass $M_g$, and Compton-$Y$ from SZE data alone. In this paper, we seek to derive the same cluster properties from the AMiBA SZE measurements of six clusters. Due to the limited $u-v$ space sampling, the AMiBA data do not provide useful constraints on the structural parameters, $\beta$ and $r_c$, in a full iterative model fitting. Instead, we adopt $\beta$ and $r_c$ from published X-ray fits and use a Markov Chain Monte-Carlo (MCMC) method to determine the cluster properties ($T_e, M_t, M_g$ and $Y$).
We also estimate these cluster properties from AMiBA data with structural constraints from X-ray data
using the non-isothermal universal temperature profile model \citep{Hallman2007}.
All quantities are integrated to spherical radius $r_{2500}$
within which the mean over-density of the cluster is $2500$~times the critical density
at the cluster's redshift.
We then investigate the scaling relations between these cluster properties derived from the SZE data,
and identify correlations between those properties 
that are induced by the iterative method.
We note that 
\citet{Locutus2008} investigate the scaling relations between the values of Compton $Y$ from AMiBA SZE data 
and other cluster properties from X-ray and other data.
All results are in good agreement.
However, we are concerned that there are embedded relations between the properties
we derived using this method. Therefore, we also investigate the embedded scaling relations
between SZE-derived properties as well.

We assume the large-scale structure of the Universe to be described by
a flat $\Lambda$CDM model with $\Omega_{\rm m} = 0.26$, $\Omega_{\rm \Lambda} =
0.74$, and Hubble constant $H_{\rm 0} = 72 \ \rm km \, s^{-1} \, Mpc^{-1}$,
corresponding to the values obtained using the WMAP 5-year data
\citep{WMAP5}. All uncertainties quoted are at the 68\%
confidence level.

\section{Determination of cluster properties}\label{sec:property}

\subsection{AMiBA Observation of SZE}\label{subsec:observation}
AMiBA is a coplanar interferometer \citep{Ho2008,Mingtang2009}. 
During 2007, it was operated with 7 close-packed antennas of 60 cm in diameter, 
giving 21 vector baselines in $u$-$v$ space and 
a synthesized resolution of $6^\prime$ \citep{Ho2008}.
The antennas are mounted on a six-meter platform \citep{Koch2008M}, 
which we rotate during the observations to provide better $u$-$v$ coverage.
The observations of SZE clusters, the details about the transform of the data into calibrated visibilities, 
and the estimated cluster profiles are presented in \citet{Wu2009}.
Further system checks are discussed in \citet{Lin2008} and \citet{amiba07-nishioka}.
For other scientific results deduced from AMiBA 2007 observations, please refer to
\citet{Locutus2008,Liu2008,Koch2008,Sandor2008,Keiichi2009}

\subsection{Isothermal $\beta$ modeling}\label{subsec:betamodel}

Because the $u$-$v$ coverage is incomplete for a single SZE experiment,
we can measure neither the accurate profile of a cluster nor its central surface brightness.
Therefore we have chosen to assume an SZE cluster model and thus a surface brightness profile,
so that a corresponding template in the $u$-$v$ space can be fitted to the observed visibilities
in order to estimate the underlying model parameters.
We consider a spherical isothermal $\beta$-model \citep{betamodel}, 
which expresses the electron number density profile as 
\begin{equation}
		n_{e}(r)=n_{{\rm e}0}\left(1+\frac{r^{2}}{r^{2}_{\rm c}}\right)^{-3\beta/2},
		\label{eq:bdensity}
\end{equation}
where $n_{{\rm e}0}$ is the central electron number density, 
$r$ is the radius from the cluster center,
$r_{\rm c}$ is the core radius, and $\beta$ is the power-law index.

Traditionally the SZE is characterized by the Compton $y$ parameter,
which is defined as the integration along the line of sight with given direction,
\begin{equation}
    y(\hat{n})\equiv\int^{\infty}_{0}\sigma_{T}n_{e}\frac{k_{\rm B}T_{\rm e}}{m_{\rm e}c^{2}}dl.
    \label{eq:def_y}
\end{equation}
Compton $y$ is related to ${\Delta}I_{\rm SZE}$ as
\begin{equation}
    {\Delta}I_{\rm SZE}=I_{\rm CMB}yf(x,T_{\rm e})\frac{xe^x}{e^x-1},
    \label{eq:intensity}
\end{equation}
where $x{\equiv}h\nu/k_{\rm B}T_{\rm CMB}$, $I_{\rm CMB}$ is the present CMB specific intensity, and
$f(x,T_{\rm e})=\left[x\coth(x/2)-4\right]\left[1+\delta_{\rm rel}(x,T_{\rm e})\right]$ \citep[e.g., ][]{L2006}.
$\delta_{\rm rel}(x,T_{\rm e})$ is a relativistic correction \citep{rel},
which we take into account to first order in $k_{\rm B}T_{\rm e}/m_{\rm e}c^{2}$. 
The relativistic correction becomes significant when the electron temperature exceeds $10~\rm keV$,
which is the regime of our cluster sample.

One can combine Equations~(\ref{eq:bdensity}-\ref{eq:intensity}) 
and integrate along the line of sight to obtain the SZE in the apparent radiation intensity as
\begin{equation}
     {\Delta}I_{\rm SZE}=I_{\rm 0}\left(1+\theta^{2}/\theta^{2}_{\rm c}\right)^{(1-3\beta)/2}
     \label{eq:betaintensity}
\end{equation}
where $\theta$ and $\theta_{\rm c}$ are the angular equivalents of $r$ and $r_{\rm c}$ respectively.
Because the clusters in our sample are not well resolved by AMiBA,
we cannot get a good estimate of $I_0$, $\beta$, and $\theta_{\rm c}$ simultaneously from our data alone.
Instead, we use the X-ray derived values for $\beta$ and $r_{\rm c}$, as summarized in Table~\ref{tab:parameters1},
and then estimate the central specific intensity $I_{\rm 0}$ \citep{Liu2008} 
by fitting Equation (\ref{eq:betaintensity}) to the calibrated visibilities obtained by \citet{Wu2009}.
In the analysis we take into account
the contamination from point sources and structures in the primary CMB.

Given the $\beta$-model described above, we can derive relations between cluster parameters and estimate them using the MCMC method.
The parameters to be estimated are 
the electron temperature $T_{\rm e}$, $r_{2500}$, total mass $M_{\rm t}\equiv M_{\rm t}(r_{2500})$, 
gas mass $M_{\rm g}\equiv M_{\rm g}(r_{2500})$,
and the integrated Compton $Y\equiv Y(r_{2500})$.

Theoretically $M_t(r_{2500})$ can be fomulated through the hydrostatic equilibrium equation
\citep[e.g., ][]{Grego2001,Bonamente2007}:
\begin{equation}
M_{\rm t}(r_{2500})=\frac{3{\beta}k_{\rm B}T_{\rm e}}{G{\mu}m_{\rm p}}\frac{r^{3}_{2500}}{r^{2}_{\rm c}+r^{2}_{2500}},
\label{eq:mtot}
\end{equation}
where $G$ is the gravitational constant and $\mu$ is the mean mass per particle
of gas in units of the mass of proton, $m_p$. 
To calculate $\mu$, we assume that
$\mu$ takes the value appropriate for clusters
with solar metallicity as given by \citet{Anders1989}.
Here we use the value $\mu=0.61$.
By combining Equation~(\ref{eq:mtot}) and the definition of $r_{2500}$, 
we can obtain $r_{2500}$ as a function of $\beta$, $T_{\rm e}$, $r_{\rm c}$, and redshift $z$ \citep[e.g., ][]{Bonamente2007}
\begin{equation}
    r_{2500}=\sqrt{\frac{3{\beta}k_{B}T_{e}}{G{\mu}m_{p}}\frac{1}{\frac{4}{3}\pi\rho_{c}(z)\cdot2500}-r^{2}_{c}}.
    \label{eq:r2500}
\end{equation}

Then $M_{\rm g}(r_{2500})$ can be expressed,
by integrating the $n_{\rm e}(r)$ in Equation~(\ref{eq:bdensity}) as
\begin{equation}
M_{g}(r)=4{\pi}\mu_{e}n_{e0}m_{p}D^{3}_{A}\int^{r/D_{A}}_{0}\left(1+\frac{\theta^{2}}{\theta^{2}_{c}}\right)^{-3\beta/2}{\theta}^{2}d\theta,
\label{eq:mgas}
\end{equation}
where $\mu_{e}=1.17$ is the mean particle mass per electron in unit of $m_{p}$, $D_{\rm A}$ is the angular diameter
determined by $z$, and $n_{{\rm e}0}$ is the central electron density, derived through the equation in \citet{L2006}:
\begin{equation}
n_{e0}=\frac{{\Delta}T_{0}m_{e}c^{2}\Gamma(\frac{3}{2}\beta)}{f(x,T_{e})T_{CMB}\sigma_{T}k_{B}T_{e}D_{A}\pi^{1/2}\Gamma(\frac{3}{2}\beta-\frac{1}{2})\theta_{c}},
\label{eq:cedensity}
\end{equation}
where $\Gamma$ is the gamma function, ${\Delta}T_0$ is the SZE temperature change, and $T_{CMB}$ is the present CMB temperature. ${\Delta}T_0$
is derived as ${\Delta}T_{0}/T_{CMB}=(e^x-1)I_{0}/xe^xI_{CMB}$.

\begin{deluxetable*}{c|cc|ccc|ccc}
\tabletypesize{\scriptsize}
\tablewidth{0pt}
\tablecaption{Parameters for isothermal spherical $\beta$-model \label{tab:parameters1}}
\tablehead{& & &\multicolumn{3}{c}{Without 100 kpc cut\tablenotemark{b}}        &  \multicolumn{3}{c}{With 100 kpc cut\tablenotemark{e}}             \\
Cluster &  z  &$D_{\rm A}$&  $\beta$  &  $r_{c}$  &  $\Delta I_{0}$\tablenotemark{c}       &   $\beta$  &  $r_{c}$  &  $\Delta I_{0}$\tablenotemark{c}          \\
        &     &  (Mpc)    &                    &   $(")$   &($\times 10^{5}$ Jy/sr)&            &   $(")$   &($\times 10^{5}$ Jy/sr)  
}
\startdata
A1689   & $0.183$ & $621$ &$0.609^{+0.005}_{-0.005}$ & $26.6^{+0.7}_{-0.7}$ & $-3.13\pm0.95$ & $0.686^{+0.010}_{-0.010}$ & 
          $48.0^{+1.5}_{-1.7}$ & $-2.36\pm0.71$ \\           
A1995   & $0.322$ & $948$ & $0.770^{+0.117}_{-0.063}$ & $38.9^{+6.9}_{-4.3}$ & $-3.30\pm1.17$ & $0.923^{+0.021}_{-0.023}$ & 
          $50.4^{+1.4}_{-1.5}$ & $-3.19\pm1.23$ \\
A2142   & $0.091$ & $340$ & $0.740^{+0.010}_{-0.010}$ & $188.4^{+13.2}_{-13.2}$ & $-2.09\pm0.36$ &          -              & 
                  -          &       -        \\
A2163   & $0.202$ & $672$ & $0.674^{+0.011}_{-0.008}$ & $87.5^{+2.5}_{-2.0}$    & $-3.24\pm0.56$ & $0.700^{+0.07}_{-0.07}$\tablenotemark{d}    & 
          $78.8^{+0.6}_{-0.6}$\tablenotemark{d} & $-3.64\pm0.61$ \\
A2261   & $0.224$ & $728$ & $0.516^{+0.014}_{-0.013}$ & $15.7^{+1.2}_{-1.1}$ & $-1.90\pm0.98$ & $0.628^{+0.030}_{-0.020}$    & 
          $29.2^{+4.8}_{-2.9}$ & $-2.59\pm0.90$ \\
A2390   & $0.232$ & $748$ & $0.600^{+0.060}_{-0.060}$\tablenotemark{a} & $28.0^{+2.8}_{-2.8}$\tablenotemark{a} & $-2.04\pm0.65$ & $0.58^{+0.058}_{-0.058}$\tablenotemark{a}    & 
          $34.4^{+3.4}_{-3.4}$\tablenotemark{a} & $-2.85\pm0.77$\\
\enddata
\tablenotetext{a} {a $10\%$ error is assumed for $\beta$ and $r_c$ for which the original reference does not give an error estimation.}
\tablenotetext{b} {Reference - \cite{2002ApJ...581...53R} for A1689, A1995, A2163, and A2261. \cite{2003MNRAS.345.1241S,2005MNRAS.359...16L} for A2142. \cite{2000MNRAS.315..269A} for A2390.}
\tablenotetext{c} {Best-fit values for $\Delta I_{0}$ with foreground estimation from point sources and CMB \citep{Liu2008}.}
\tablenotetext{d} {$\beta$ fixed to a fiducial value $0.7$ in \cite{2006ApJ...647...25B}, a $10\%$ error is assumed.} 
\tablenotetext{e} {Reference - \cite{2006ApJ...647...25B} for A1689, A1995, A2163, and A2261. \cite{2001MNRAS.324..877A} for A2390.}
\end{deluxetable*}

Finally, with the $I_0$ computed earlier and the $r_{2500}$ estimated here 
we can integrate the Compton $y$ out to $r_{2500}$ to yield $Y$
\begin{equation}
Y=\frac{2{\pi}{\Delta}T_{0}}{f(x,T_{e})T_{CMB}}\int^{\theta_{2500}}_{0}\left(1+\frac{\theta^{2}}{\theta^{2}_{c}}\right)^{(1-3\beta)/2}\theta d\theta,
\label{eq:intY}
\end{equation}
where $\theta_{2500}=r_{2500}/D_{a}$ indicates the projected angular size of $r_{2500}$.

With the formulae as described above, for a set of $\beta$, $r_{\rm c}$, and $z$ as measured from X-ray observations and $I_{0}$ from AMiBA SZE observation, we can arbitrarily assign a `pseudo' electron temperature $T_{{\rm e}(i)}$,
and then determine the pseudo $r_{2500}(T_{{\rm e}(i)})$, $M_{\rm t}(T_{{\rm e}(i)})$, $M_{\rm g}(T_{{\rm e}(i)})$, and $Y(T_{{\rm e}(i)})$. Given $M_{\rm t}(T_{{\rm e}(i)})$ and $M_{\rm g}(T_{{\rm e}(i)})$,
we obtained the pseudo gas fraction $f_{gas}(T_{{\rm e}(i)})=M_{\rm g}(T_{{\rm e}(i)})/M_{\rm t}(T_{{\rm e}(i)})$.
Using $f_{gas}(T_{{\rm e}(i)})$ as a function of $T_{{\rm e}(i)}$ we applied the MCMC method by varying $T_{\rm e}$ and $\Delta I_{0}$ to estimate the likelihood distribution
of each cluster property. While estimating the MCMC likelihood we assume that the likelihoods
of $\Delta I_{0}$ and $f_{gas}$ are independent. The likelihood distributions of $\Delta I_{0}$ for each cluster are 
taken from the fitting results of \citet{Liu2008}, while the likelihood distribution of $f_{gas}$ is
assumed to be Gaussian with mean $0.116$ and standard deviation $0.005$, which is the ensemble average over 38 clusters observed by Chandra and OVRO/BIMA \citep{L2006}.

\begin{figure*}
\plotone{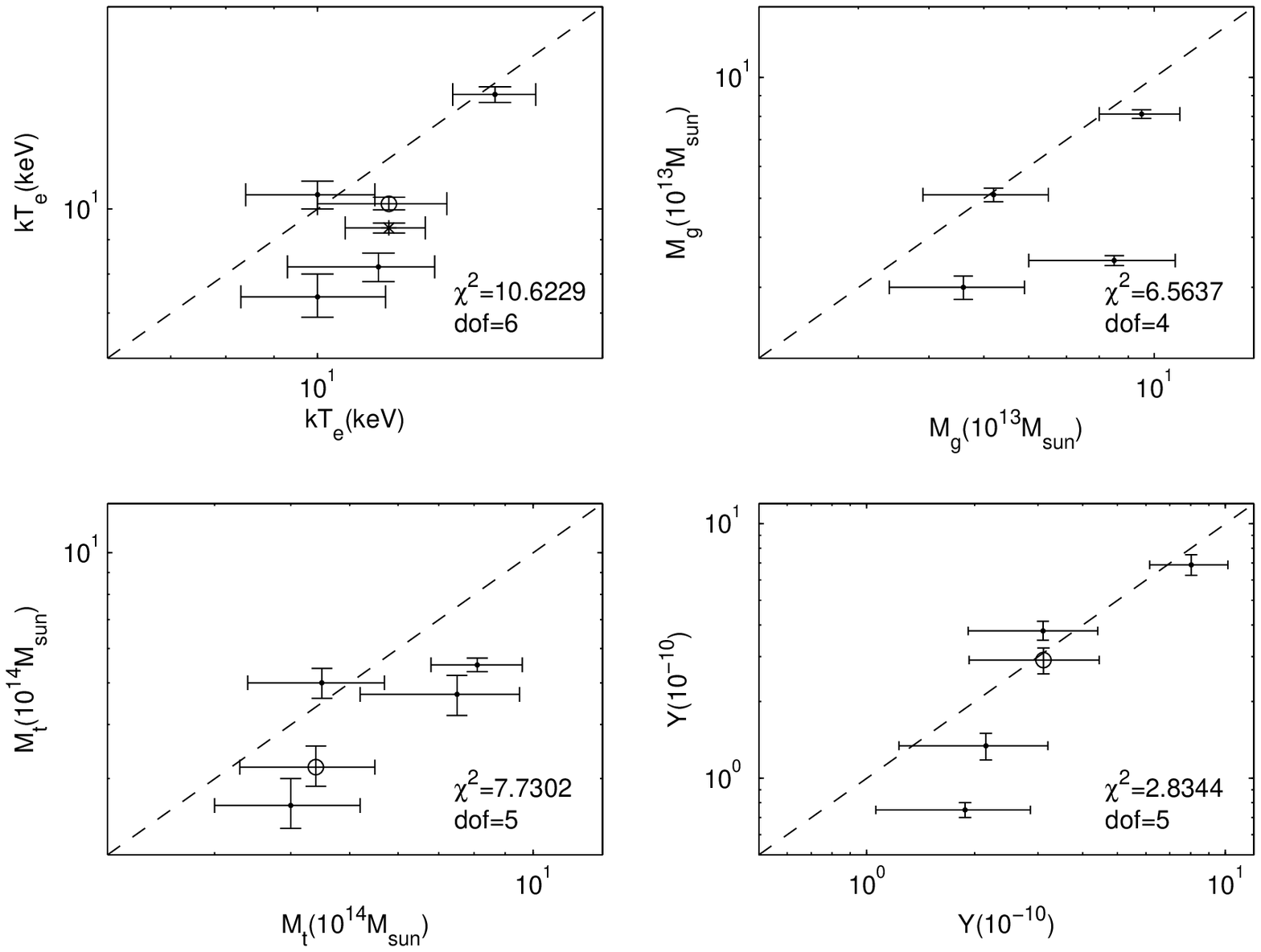}
\caption{\small 
Comparison of $T_{\rm e}$ (upper-left), $M_{\rm g}$ (upper-right), $M_{\rm t}$ (lower-left),
and $Y$ (lower-right) of clusters derived from AMiBA SZE data based on isothermal $\beta$ model with 
100-kpc cut (x-axis) and those given in literature (y-axis). 
All y-axis values are from \citet{Bonamente2007},
except for
the Y values, which are from \citet{Morandi2007},
and those for A2390, which is indicated by a circle
with $T_e$ from \citet{Benson2004} and $M_{\rm t}$ calculated from the data in \citet{Benson2004}. 
The dashed lines indicate $y~=~x$.
\label{fig:com}}
\end{figure*}

In the process, the values of $\beta$, $r_{\rm c}$, and $z$
are taken from other observational results which are summarized in \citet{Koch2008} and Table~\ref{tab:parameters1}. We took the $\beta$ model parameters from both ROSAT and Chandra X-ray results.
The Chandra results were derived by fitting an isothermal $\beta$ model to
the X-ray data with a central 100-kpc cut. The aim of the cut-off is to exclude
the complicated non-gravitational physics (e.g, radiative
cooling and feedback mechanisms) in cluster cores.
Table ~\ref{tab:iso} summarizes our results derived assuming an isothermal $\beta$ model.
We present the results obtained with isothermal $\beta$ model parameters derived with and without 100-kpc cut
both here.
Figure~\ref{fig:com} compares our results with the SZE-X-ray joint results obtained from OVRO/BIMA and Chandra data
\citep{Bonamente2007,Morandi2007}.
These are in good agreement.

\begin{deluxetable*}{c|ccccc|ccccc}
\tabletypesize{\scriptsize}
\tablewidth{0pt}
\tablecaption{SZE derived cluster properties in isothermal $\beta$ model\label{tab:iso}}
\tablehead{
 &\multicolumn{5}{c}{Without 100-kpc cut}&\multicolumn{5}{c}{With 100-kpc cut}\\
Cluster  & $r_{2500}$ & $k_{\rm B}T_{\rm e}$ & $M_{\rm g}$ & $M_{\rm t}$ & $Y$ & $r_{2500}$ & $k_{\rm B}T_{\rm e}$ & $M_{\rm g}$ & $M_{\rm t}$ & $Y$ \\
 & $(")$ & (keV) & $(10^{13}M_{\odot})$ & $(10^{14}M_{\odot})$ & $(10^{-10})$ &
 $(")$ & (keV) & $(10^{13}M_{\odot})$ & $(10^{14}M_{\odot})$ & $(10^{-10})$ 
}
\startdata
A1689 &$209^{+16}_{-19}$&$10.4^{+1.6}_{-1.7}$&$4.9^{+1.2}_{-1.2}$&$4.2^{+1.1}_{-1.0}$&$3.2^{+1.5}_{-1.2}$
&$214^{+16}_{-19}$&$10.0^{+1.5}_{-1.6}$&$5.2^{+1.3}_{-1.3}$&$4.5^{+1.2}_{-1.1}$&$3.1^{+1.3}_{-1.2}$\\
A1995 &$150^{+13}_{-15}$&$12.0^{+1.9}_{-2.2}$&$7.4^{+1.9}_{-2.1}$&$6.4^{+1.7}_{-1.8}$&$1.9^{+1.0}_{-0.8}$
&$159^{+13}_{-18}$&$11.6^{+1.7}_{-2.3}$&$8.5^{+2.4}_{-2.5}$&$7.5^{+2.0}_{-2.3}$&$1.9^{+1.0}_{-0.8}$\\
A2142 &$430^{+23}_{-28}$&$11.9^{+1.1}_{-1.3}$&$6.6^{+1.1}_{-1.2}$&$5.7^{+1.0}_{-1.0}$&$16.9^{+4.4}_{-4.2}$
&$     -     $&$     -     $&$     -     $&$     -     $&$     -     $\\
A2163 &$228^{+14}_{-13}$&$15.3^{+1.7}_{-1.5}$&$8.5^{+1.5}_{-1.5}$&$7.2^{+1.4}_{-1.2}$&$7.7^{+2.2}_{-1.9}$
&$237^{+13}_{-13}$&$15.4^{+1.6}_{-1.5}$&$9.5^{+1.6}_{-1.5}$&$8.1^{+1.5}_{-1.3}$&$8.0^{+2.1}_{-1.9}$\\
A2261 &$147^{+15}_{-20}$&$8.7^{+1.8}_{-2.3}$&$2.7^{+1.0}_{-1.0}$&$2.3^{+0.9}_{-0.9}$&$1.3^{+1.0}_{-0.8}$
&$172^{+16}_{-15}$&$10.0^{+1.8}_{-1.7}$&$4.6^{+1.3}_{-1.2}$&$4.0^{+1.2}_{-1.0}$&$2.2^{+1.1}_{-0.9}$\\
A2390 &$156^{+12}_{-15}$&$9.2^{+1.3}_{-1.7}$&$3.7^{+0.9}_{-1.0}$&$3.2^{+0.8}_{-0.8}$&$1.6^{+0.7}_{-0.6}$
&$174^{+13}_{-15}$&$11.9^{+1.8}_{-1.9}$&$5.2^{+1.3}_{-1.2}$&$4.4^{+1.1}_{-1.1}$&$3.1^{+1.3}_{-1.2}$\\
\enddata
\end{deluxetable*}
\subsection{UTP $\beta$ model}\label{subsec:UTP}

The simulation done by \citet{Hallman2007} suggested incompatibility between isothermal $\beta$
model parameters fitted to X-ray surface brightness profiles and those fitted to SZE profiles. This incompatibility
also causes bias in the estimates of $Y$ and $M_{g}$. They suggested a non-isothermal $\beta$ model
with a universal temperature profile (UTP). We also considered how the UTP $\beta$ model changes our
estimates of cluster properties in this section.

In the UTP $\beta$-model, the baryon density profile is the same as Equation~ (\ref{eq:bdensity}), and the temperature profile can be written as \citep{Hallman2007}:
\begin{equation}
T_{e}\left(r\right)=\left\langle T\right\rangle_{500}T_{0}\left(1+\left(\frac{r}{\alpha r_{500}}\right)^{2}\right)^{-\delta},
\label{eq:tprofile}
\end{equation}
where $\left\langle T\right\rangle_{500}$ indicates the average spectral temperature inside $r_{500}$. $T_{0}$, $\alpha$, and $\delta$ are dimensionless parameters in the universal temperature profile model.
$\delta$ is the outer slope of the temperature profile, outside of a
 core with electron temperature $T_{\rm e0}=\left\langle T\right\rangle_{500}T_{0}$.
 This core is of size $\alpha r_{500}$.
The total mass can be obtained by solving the hydrostatic equilibrium equation \citep{Fabricant1980}:
\begin{equation}
M_{t}\left(r\right)=-\frac{k_{\rm B}r^{2}}{G\mu m_{\rm p}}\left(T_{\rm e}(r)\frac{dn_{\rm e}(r)}{dr}+n_{\rm e}(r)\frac{dT_{\rm e}(r)}{dr}\right).
\label{eq:hydrostatic}
\end{equation}
In the isothermal $\beta$-model, Equation~(\ref{eq:hydrostatic}) can be reduced into the form of Equation~(\ref{eq:mtot}). However,
in the UTP $\beta$-model, the derivative of $T_{\rm e}(r)$ with respect to $r$ in Equation~(\ref{eq:hydrostatic}) is no longer zero. By applying
Equation~(\ref{eq:bdensity}) and Equation~(\ref{eq:tprofile}) in Equation~(\ref{eq:hydrostatic}), one can obtain:
\begin{equation}
M_{t}\left(r\right)=\frac{k_{\rm B}T_{\rm e0}}{G\mu m_{\rm p}}\left(\frac{3\beta r^{3}}{r^{2}+r^{2}_{\rm c}}+\frac{2\delta r^{3}}{r^{2}+\alpha^{2}r^{2}_{500}}\right)\left(1+\frac{r^{2}}{\alpha^{2}r^{2}_{500}}\right)^{-\delta}.
\label{eq:mtotutp}
\end{equation}

By combining Equation~(\ref{eq:mtotutp}) and the definition of $r_{500}$, an analytical solution for $r_{500}$ can be obtained as:
\begin{equation}
r_{500}=\sqrt{\frac{(1+\alpha^{2})(3\beta A-r^{2}_{c})+2\delta A+\sqrt{D}}{2(1+\alpha^{2})}},
\label{eq:r500utp}
\end{equation}
where $A=3k_{\rm B}T_{\rm e0}(1+\alpha^{-2})^{-\delta}/(4G\mu m_{\rm p}\pi\rho_{\rm c}(z)\cdot 500)$, and
$D=[(1+\alpha^{2})(3\beta A-r^{2}_{c})+2\delta A]^{2}+8(1+\alpha^{2})\delta Ar^{2}_{\rm c}$. If $\delta \rightarrow 0$ or 
$\alpha \rightarrow \infty$, which indicate the nearly isothermal case, Equation~(\ref{eq:r500utp}) reduces to a form similar to Equation~(\ref{eq:r2500}). 

Using the definition of $r_{500}$, $M_{t}(r_{500})$ can be written as:
\begin{equation}
M_{t}(r_{500})=500\cdot\frac{4}{3}\pi r^{3}_{500}\rho_{\rm c}(z).
\label{eq:mt500}
\end{equation}

For an arbitrary overdensity $\Delta$, we can not find an analytical solution for arbitrary $r_{\Delta}$ (i.e.: $r_{2500}$, $r_{200}$, etc.). However, with the known $r_{500}$, we can still find the numerical solution for $r_{\Delta}$ easily. We can then solve for $M_{t}(r_{\Delta})$ using Equation~(\ref{eq:mtotutp}). 

To yield the central electron number density, we consider the formula for the Compton $y$ resulting from the UTP $\beta$-model (see the Appendix of \citet{Hallman2007}). By setting the projected radius $b=0$ in Equation~(A10) in \citet{Hallman2007}, one can obtain:
\begin{equation}
n_{e0}=\frac{{\Delta}T_{0}m_{e}c^{2}}{f(x,T_{e})T_{CMB}\sigma_{T}k_{B}\left\langle T\right\rangle_{500}T_{0}I_{\rm SZ}(0)},
\label{eq:ne0utp}
\end{equation}
where
\begin{equation}
I_{\rm SZ}(0)=\frac{\pi^{1/2}\Gamma(\frac{3}{2}\beta+\delta-\frac{1}{2})F_{2,1}\left(\delta ,\frac{1}{2};\frac{3\beta}{2}+\delta,1-\frac{r^{2}_{\rm c}}{\alpha^{2}r^{2}_{500}}\right)r_{c}}{\Gamma(\frac{3\beta}{2}+\delta)},
\label{eq;isz0}
\end{equation}
and $F_{2,1}$ is Gauss' hypergeometric function. Here we assume $f(x,T_{e})=f(x,\left\langle T\right\rangle_{500}T_{0})$, and the change of $f(x,T_{e})$ due to the change of $T_{e}$ along line of sight
is negligible. Actually, by numerical calculation we found that the error in Equation~(\ref{eq:ne0utp}) caused by this assumption is less than $~1\%$. Because the UTP $\beta$ model assumes the electron density profile as same as the isothermal
$\beta$ model, we can rewrite $M_{g}$ in UTP model by simply applying Equation (\ref{eq:ne0utp}) in Equation (\ref{eq:mgas}).

Thus, the integration of the Compton $y$ profile, instead of Equation~(\ref{eq:intY}), becomes:
\begin{equation}
Y=Y_{0}\int^{\theta_{2500}}_{0}\left(1+\frac{\theta^{2}}{\theta^{2}_{c}}\right)^{(1-3\beta)/2}\left(1+\frac{\theta^{2}}{\alpha^{2}\theta^{2}_{500}}\right)^{-\delta}F(\theta)\theta d\theta,
\label{eq:intYutp}
\end{equation}
where $Y_{0}=(2{\pi}{\Delta}T_{0})/(fT_{\rm CMB}F(0))$ and 
$F(\theta)=F_{2,1}(\delta ,1/2;3\beta/2+\delta,1-(r^{2}_{\rm c}+\theta^{2})/(\alpha^{2}r^{2}_{500}+\theta^{2}))$.

We were not able to constrain the parameters $\beta$, $r_{c}$, $\delta$, and $\alpha$ of the UTP significantly with our SZE data alone. However, the simulation of \citet{Hallman2007} suggested that there is no significant systematic difference
between the values of $\beta$ and $r_{c}$ resulting from fitting an isothermal $\beta$ model to mock X-ray observations
and those parameters fitted using the UTP $\beta$ model. Therefore, we simply assume that the ratio between the isothermal $\beta_{\rm iso}$ value and UTP $\beta_{\rm UTP}$ value is $1\pm0.1$, and $r_{c,\rm iso}/r_{c,\rm UTP}=1\pm0.2$, for each cluster.
We also assume $\delta=0.5$, $\alpha=1$, and $T_{0}=1.3$. Those values are taken from the average of results of \citet{Hallman2007}. Then we fit $\Delta I_{0}$ to AMiBA SZE observation data with the UTP $\beta$ model parameters above by fixing $\delta$, $\alpha$, and $T_{0}$, and treating the likelihood distributions of $\beta_{\rm UTP}$ and $r_{c,\rm UTP}$ as two independent Gaussian-distributions. 
Finally, we applied the MCMC method, which varies $\Delta I_{0}$, $\beta$, $r_{c}$, and $\left\langle T\right\rangle_{500}$, to estimate cluster properties with the equations derived from the UTP $\beta$ model and the data fitting results. 

\begin{figure*}
	\plotone{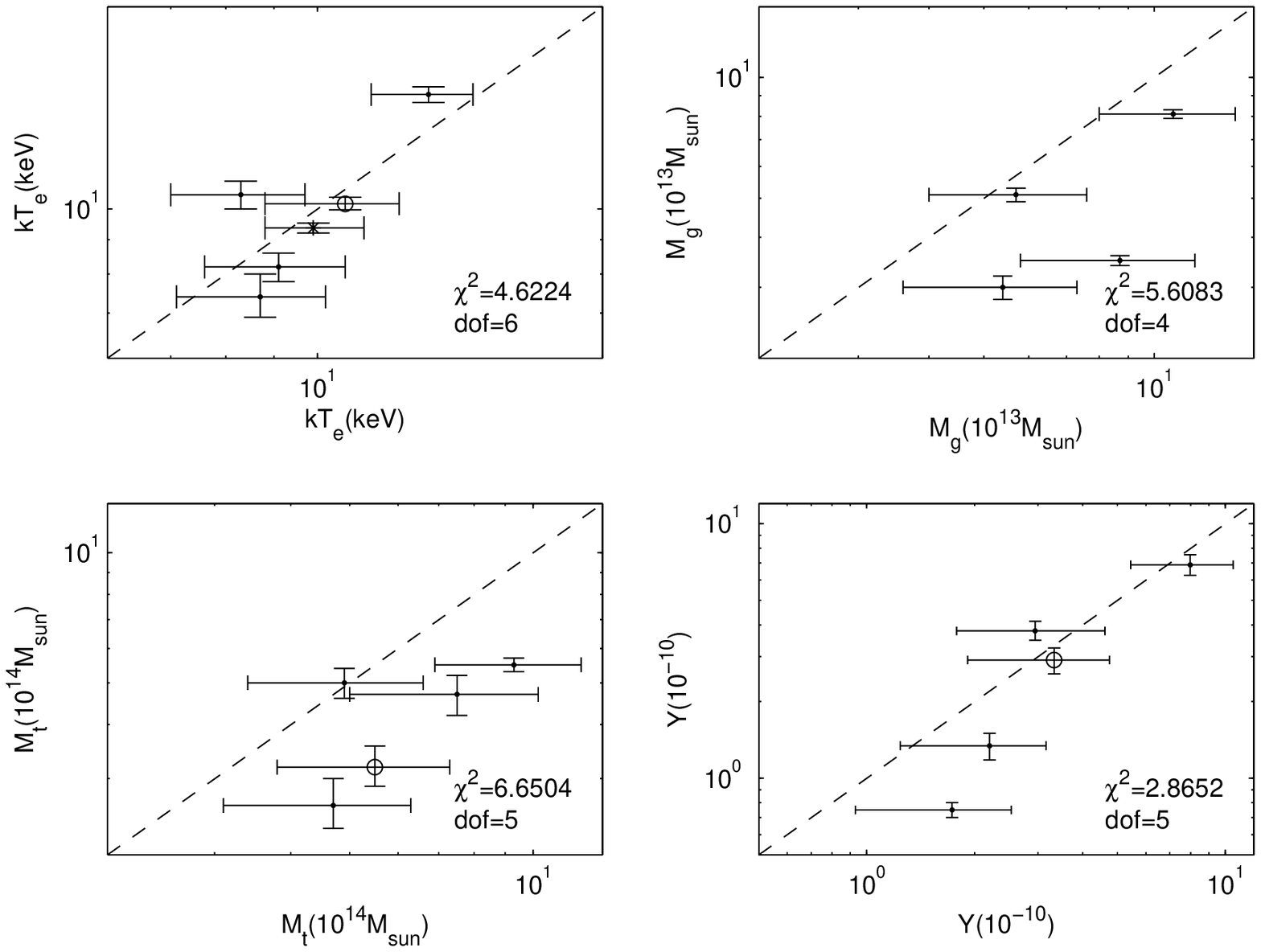}
  \caption{\small 
Comparison of $T_{\rm e}$ (upper-left), $M_{\rm g}$ (upper-right), $M_{\rm t}$ (lower-left),
and $Y$ (lower-right) of clusters derived from AMiBA SZE data based on the UTP $\beta$ model with 
100-kpc cut (x-axis) and those given in literature (y-axis). 
All y-axis values are from \citet{Bonamente2007},
except for
the Y values, which are from \citet{Morandi2007},
and those for A2390, which is indicated by a circle
with $T_e$ from \citet{Benson2004} and $M_{\rm t}$ calculated from the data in \citet{Benson2004}. 
The dashed lines indicate $y~=~x$.
\label{fig:comutp}}
\end{figure*}

Table ~\ref{tab:utp} summarizes our results derived with the UTP $\beta$ model.
Figure~\ref{fig:comutp} compares our results with the SZE-X-ray joint results obtained from OVRO/BIMA and Chandra data
\citep{Bonamente2007,Morandi2007}.
These are also in good agreement. We find that the electron temperature derived with
the UTP $\beta$ model are in significantly better agreement with the temperatures from Chandra X-ray measurements.


\begin{deluxetable*}{c|ccccc|ccccc}
\tabletypesize{\scriptsize}
\tablewidth{0pt}
\tablecaption{SZE derived cluster properties in the UTP $\beta$ model\label{tab:utp}}
\tablehead{
 &\multicolumn{5}{c}{Without 100-kpc cut}&\multicolumn{5}{c}{With 100-kpc cut}\\
Cluster  & $r_{2500}$ & $k_{\rm B}T_{\rm e}$\tablenotemark{a} & $M_{\rm g}$ & $M_{\rm t}$ & $Y$ & $r_{2500}$ & $k_{\rm B}T_{\rm e}$\tablenotemark{a} & $M_{\rm g}$ & $M_{\rm t}$ & $Y$ \\
 & $(")$ & (keV) & $(10^{13}M_{\odot})$ & $(10^{14}M_{\odot})$ & $(10^{-10})$ &
 $(")$ & (keV) & $(10^{13}M_{\odot})$ & $(10^{14}M_{\odot})$ & $(10^{-10})$ 
}
\startdata
A1689 &$219^{+23}_{-23}$&$8.9^{+1.5}_{-1.6}$&$5.6^{+1.9}_{-1.8}$&$4.8^{+1.7}_{-1.5}$&$3.4^{+1.7}_{-1.5}$
&$220^{+23}_{-22}$&$8.3^{+1.4}_{-1.3}$&$5.7^{+1.9}_{-1.7}$&$4.9^{+1.7}_{-1.5}$&$3.0^{+1.5}_{-1.2}$\\
A1995 &$154^{+16}_{-18}$&$9.7^{+1.7}_{-1.7}$&$7.6^{+2.8}_{-2.5}$&$6.7^{+2.3}_{-2.3}$&$1.8^{+1.1}_{-0.8}$
&$161^{+16}_{-20}$&$9.1^{+1.6}_{-1.5}$&$8.7^{+3.1}_{-2.9}$&$7.5^{+2.7}_{-2.5}$&$1.9^{+1.0}_{-0.8}$\\
A2142 &$458^{+43}_{-49}$&$9.9^{+1.1}_{-1.3}$&$7.6^{+2.4}_{-2.3}$&$6.4^{+2.2}_{-1.8}$&$17.0^{+6.5}_{-5.5}$
&$     -     $&$     -     $&$     -     $&$     -     $&$     -     $\\
A2163 &$245^{+23}_{-23}$&$13.0^{+1.7}_{-1.5}$&$10.1^{+3.0}_{-2.7}$&$8.8^{+2.6}_{-2.4}$&$8.0^{+3.1}_{-2.5}$
&$251^{+21}_{-24}$&$13.1^{+1.5}_{-1.7}$&$10.8^{+3.1}_{-2.8}$&$9.3^{+2.7}_{-2.4}$&$8.3^{+2.9}_{-2.6}$\\
A2261 &$160^{+17}_{-22}$&$7.9^{+1.5}_{-1.8}$&$3.6^{+1.3}_{-1.4}$&$3.1^{+1.1}_{-1.2}$&$1.5^{+0.9}_{-0.8}$
&$183^{+20}_{-21}$&$8.7^{+1.5}_{-1.6}$&$5.4^{+1.9}_{-1.8}$&$4.7^{+1.6}_{-1.6}$&$2.2^{+1.2}_{-1.0}$\\
A2390 &$166^{+17}_{-17}$&$8.1^{+1.2}_{-1.4}$&$4.5^{+1.5}_{-1.4}$&$3.9^{+1.3}_{-1.2}$&$1.8^{+0.8}_{-0.7}$
&$188^{+18}_{-20}$&$10.7^{+1.5}_{-1.9}$&$6.3^{+2.1}_{-1.9}$&$5.5^{+1.8}_{-1.7}$&$3.3^{+1.6}_{-1.4}$\\
\enddata
\tablenotetext{a}{The average electron temperature up to $r_{500}$ (i.e.:$\left\langle T\right\rangle_{500}$ in Equation (\ref{eq:tprofile}))}
\end{deluxetable*}

\section{Embedded scaling relations}\label{sec:scaling}

The self-similar model \citep{Kasier1986} predicts simple power-law scaling relations between
cluster properties \citep[e.g., ][]{Bonamente2007,Morandi2007}.
Motivated by this, people usually investigate the scaling relations between the derived cluster
properties from observational data to see whether they are consistent with the self-similar model.
However, the method described above is based on the isothermal $\beta$-model and the UTP $\beta$-model. Therefore, there could be
some embedded relations which agree with self-similar model predictions between the derived properties. We investigated the embedded relations through both
analytical and numerical methods.

\subsection{Analytical formalism and numerical analysis}\label{sec:ana}

In the isothermal $\beta$ model, by applying Equation~(\ref{eq:r2500}) in Equation~(\ref{eq:mtot}), $M_{\rm t}$ can be rewritten
as
\begin{equation}
    M_{\rm t}=2500\cdot\frac{4}{3}\pi\rho_{\rm c}\left(z\right)\left(\frac{3{\beta}k_{\rm B}T_{\rm e}}{G{\mu}m_{\rm p}}\frac{1}{2500\cdot\frac{4}{3}\pi\rho_{\rm c}\left(z\right)}-r^{2}_{\rm c}\right)^{\frac{3}{2}}.
    \label{eq:mtrelation}
\end{equation}
As we can see, while $\beta$ is set to be a constant, and $r^{2}_{2500}>>r^{2}_{\rm c}$, which implies $3{\beta}k_{\rm B}T_{\rm e}/(G{\mu}m_{\rm p}\cdot2500\cdot\frac{4}{3}\pi\rho_{\rm c}\left(z\right))>>r^{2}_{\rm c}$, the relation
$M_{t}{\propto}T^{3/2}_{\rm e}$ will be obtained. 
However, for some of the clusters we considered in this paper, the values of $r_{2500}/r_{\rm c}$ are only slightly above $2$. Therefore,
we have to investigate the scaling relation between $M_{\rm t}$ and $T_{\rm e}$ by considering $\partial \ln M_{\rm t}/\partial \ln T_{\rm e}$.

By partially differentiating Equation (\ref{eq:mtrelation}) by $T_{\rm e}$, and multiplying it by $T_{\rm e}/M_{\rm t}$, we can get
\begin{equation}
      \frac{\partial \ln M_{\rm t}}{\partial \ln T_{\rm e}}=\frac{3}{2}\frac{(r^{2}_{2500}+r^{2}_{\rm c})}{r^{2}_{2500}},
      \label{eq:slopeMT}
\end{equation}
which decreases from $1.875$ at $r_{2500}/r_{\rm c}=2$ to $1.5$ as $r_{2500}/r_{\rm c}\rightarrow \infty$. That implies $M_{\rm t}$ behaves as $M_{\rm t}\propto T^{1.875}_{\rm e}$ while $r_{2500}/r_{\rm c}\approx 2$ and $M_{\rm t}\propto T^{1.5}_{\rm e}$ while $r_{2500}/r_{\rm c}$ approaches infinity. 
This result shows that there is an embedded $M_{\rm t}$-$T_{\rm e}$ relation consistent
with the self-similar model in the method described above.

If we assume that the gas fraction $f_{\rm gas}$ is a constant, the
scaling relation between $M_{\rm g}$ and $T_{\rm e}$ will be as same as the relation between $M_{\rm t}$ and $T_{\rm e}$.

In order to investigate the relations between integrated $Y$ and the other cluster properties, we consider Equation~(\ref{eq:intY}). By combining Equation~(\ref{eq:r2500})-(\ref{eq:cedensity}), one can obtain:
\begin{equation}
    {\Delta}T_{0}=\frac{M_{\rm g}(r_{2500})f(x,T_{e})T_{\rm CMB}\sigma_{\rm T}k_{\rm B}T_{\rm e}\Gamma\left(\frac{3}{2}\beta-\frac{1}{2}\right)\theta_{\rm c}}{4\pi^{1/2}\mu_{\rm e}m_{\rm p}D^{2}_{\rm A}m_{\rm e}c^{2}\Gamma\left(\frac{3}{2}\beta\right)\int^{\theta_{2500}}_{0}\left(1+\frac{\theta^{2}}{\theta^{2}_{\rm c}}\right)^{-3\beta/2}\theta^{2}d\theta}.
    \label{eq:deltaT0}
\end{equation}
Then we combine Equation (\ref{eq:deltaT0}) and Equation (\ref{eq:intY}) and obtain:
\begin{equation}
     Y=\frac{\pi^{1/2}M_{\rm g}(r_{2500})\sigma_{\rm T}k_{\rm B}T_{\rm e}}{2\mu_{\rm e}m_{\rm p}m_{\rm e}c^{2}D^{2}_{\rm A}}g(\theta_{2500},\theta_{\rm c},\beta),
     \label{eq:ytrelation}
\end{equation}
where
\begin{equation}
g(\theta_{2500},\theta_{\rm c},\beta)=\frac{\Gamma\left(\frac{3}{2}\beta-\frac{1}{2}\right)\theta_{\rm c}\int^{\theta_{2500}}_{0}\left(1+\frac{\theta^{2}}{\theta^{2}_{\rm c}}\right)^{\left(1-3\beta\right)/2}\theta d\theta}{\Gamma\left(\frac{3}{2}\beta\right)\int^{\theta_{2500}}_{0}\left(1+\frac{\theta^{2}}{\theta^{2}_{\rm c}}\right)^{-3\beta/2}\theta^{2}d\theta}
     \label{eq:funG}
\end{equation}
is a dimensionless function of $\theta_{2500}$, $\theta_{\rm c}$, and $\beta$. 

\begin{figure}
	\plotone{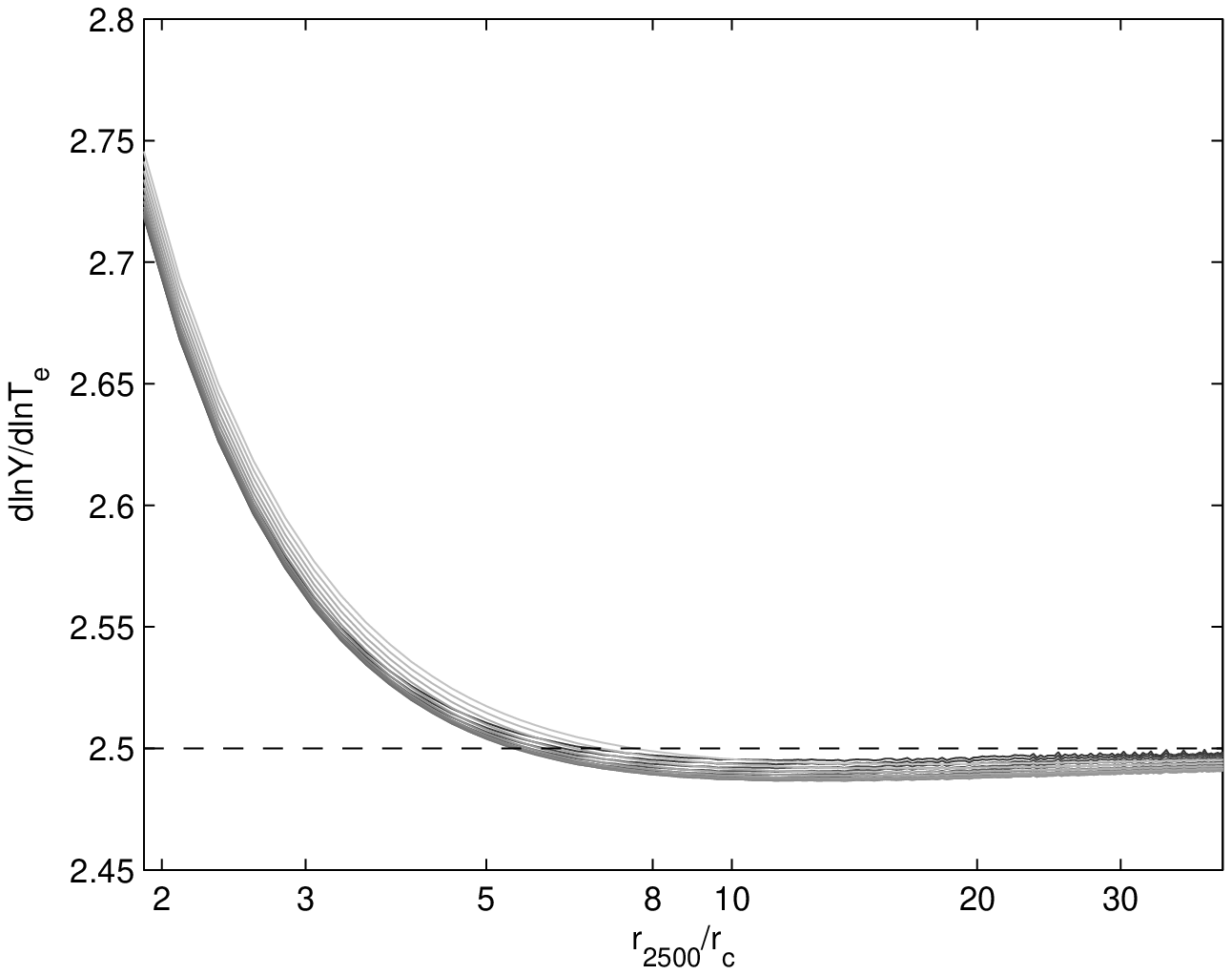}
  \caption{\small 
Embedded scaling relation between $Y$ and $T_{\rm e}$. The shaded scale indicates different $\beta$
from $0.5$ (the darkest line) to $1.2$ (the lightest line). The dashed line indicates the predicted value by
self-similar model.
\label{fig:YTembed}}
\end{figure}

We also calculated $\partial \ln Y/\partial \ln T_{\rm e}$ to investigate the behavior of $Y$ when $T_{\rm e}$ varies (see Figure~\ref{fig:YTembed}). As we can see in Figure~\ref{fig:YTembed}, $\partial \ln Y/\partial \ln T_{\rm e}$ varies between $2.45$ and $2.75$ while $r_{2500}/r_{\rm c}>2.0$ and $0.5\leq\beta\leq1.2$. We also noticed that
$\partial \ln Y/\partial \ln T_{\rm e}$ approaches $2.5$ as $r_{2500}/r_{c}$ approaches infinity.
This result indicates that behaviour similar to the self-similar model is built into 
scaling relation studies based solely on SZE data.

The effect of varying $\beta$ is investigated. If we consider power law scaling
relation
\begin{equation}
Q=10^{A}X^{B}
\label{eqn:line}
\end{equation}
between $M_{\rm t}$ and $T_{\rm e}$ with $M_{\rm t}$ written as Equation (\ref{eq:mtrelation}), one can find that changing the value of $\beta$ will only affect the normalization factor $A$. In other words, if we change $\beta$ to $\beta'$, $A$ will be changed to $A'=A+B\log_{10}(\beta'/\beta)$.

In the $Y$-$T_{e}$ relation, $\beta$ will affect the scaling power $B$ as shown in Figure \ref{fig:YTembed}.
$B$ varies within a range of only $0.04$ while $0.5\leq\beta\leq1.2$.

\begin{figure}
	\plotone{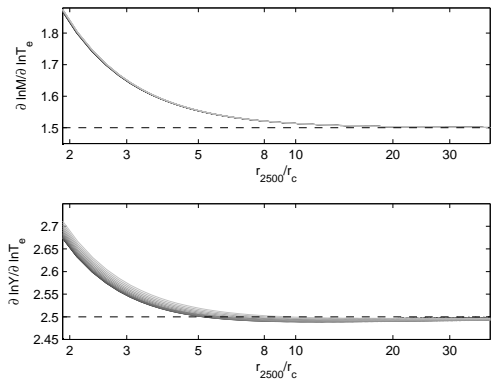}
  \caption{\small 
Embedded $M_{t}$-$T_{\rm e}$ (upper panel) and $Y$-$T_{\rm e}$ (lower panel) scaling relations in UTP $\beta$
model. The grey scales indicate different $\beta$
from $0.5$ (the darkest line) to $1.2$ (the lightest line). The dashed lines indicate the predicted values by
self-similar model.
\label{fig:utpembed}}
\end{figure}

Considering the UTP $\beta$ model, we undertook a similar analysis of the embedded scaling relation. The results, which are similar with those obtained with the isothermal $\beta$ model, are shown in Figure \ref{fig:utpembed}.

\subsection{Calculation of Scaling Relations}
Here we investigate the $Y$-$T_{\rm e}$, $Y$-$M_{\rm t}$, and $Y$-$M_{\rm g}$ scaling relations
for the quantities derived above.
We also study the $M_{\rm t}$-$T_{\rm e}$ scaling relation 
with the $M_{\rm t}$ from AMiBA SZE data and the $T_{\rm e}$ from X-ray data \citep{Bonamente2007,Morandi2007}.

\begin{figure}
\epsscale{0.8}
\plotone{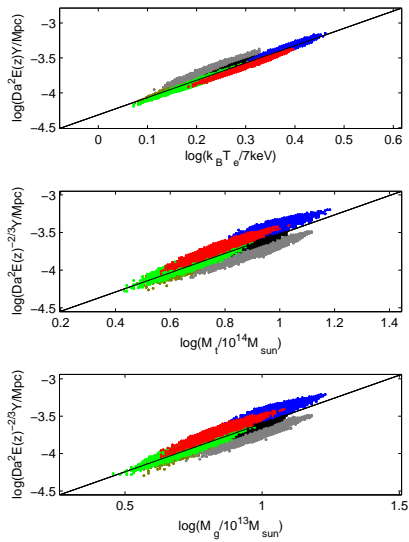}
  \caption{\small 
Scaling Relations of 
$Y$-$T_{\rm e}$ (upper), $Y$-$M_{\rm g}$ (middle), and $Y$-$M_{\rm t}$ (lower)
based on the AMiBA SZE derived results. 
Gray areas indicate the 68\% confidence regions for the parameter pairs of each cluster.
Solid lines are the best fits as in Tab.~\ref{tab:sr}.\label{fig:Ysr}}
\end{figure}

\begin{figure}
\plotone{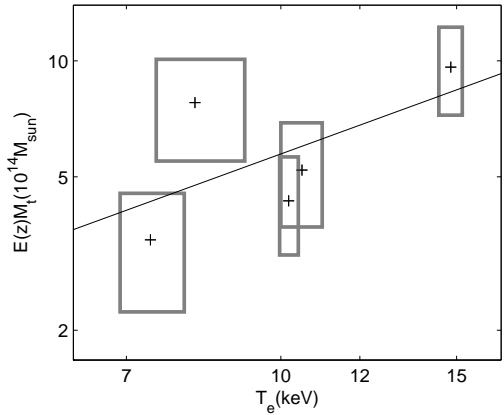}
\caption{\small
$M_{\rm t}-T_{\rm e}$ scaling relation
between the X-ray measured $T_{\rm e}$ \citep{Bonamente2007,Morandi2007}
and the AMiBA derived $M_{\rm t}$. 
The boxes indicate the $1\sigma$ errors for each cluster.
The solid line is the best fit as in Tab.~\ref{tab:sr}.
 \label{fig:MTsr}}
\end{figure}

For a pair of cluster properties $Q$-$X$, we consider the power-law scaling relation (Equation (\ref{eqn:line})).
To estimate $A$ and $B$,
we perform a maximum-likelihood analysis in the log-log plane.
For the $M_{\rm t}$-$T_{\rm e}$ relation,
because $M_{\rm t}$ and $T_{\rm e}$ are independent measurements from different observational data,
we can simply perform linear minimum-$\chi^{2}$ analysis to estimate $A$ and $B$ \citep{Press1992,Benson2004}.
On the other hand, for the SZE-derived properties, 
because they are correlated and so are their likelihoods 
(i.e., $L(Q,X)\neq L(Q)L(X)$, as manifested by the colored areas in Figure~\ref{fig:Ysr}),
we cannot apply $\chi^{2}$ analysis.
Instead we use a Monte Carlo method by randomly choosing one MCMC iteration from each cluster many times.
With each set of iterations we derived a pair of $A_{i}$ and $B_{i}$ using linear regression method.
Finally we estimate the likelihood distribution of $A$ and $B$ using the distribution of $\left\{A_{i}\right\}$ and
$\left\{B_{i}\right\}$.
The results are presented in Table~\ref{tab:sr} and Figures~\ref{fig:Ysr} and~\ref{fig:MTsr}.
However, as we discussed in Section \ref{sec:ana}, the scaling relations between
SZE-derived properties should be interpreted as a test of embedded scaling relations rather than estimations of the true scaling relations. On the other hand, the $M_{\rm t}$-$T_{\rm e}$ relation
compared $M_{\rm t}$ and $T_{\rm e}$ from different experiments. Therefore, we can regard it 
as a test of the scaling relation prediction.

\begin{deluxetable}{c|ccc}
\tabletypesize{\scriptsize}
\tablecaption{Scaling relations of SZE-derived cluster properties\label{tab:sr}}
\tablehead{
Scaling& & & \\
Relations& $A$ & $B$ & $B_{\rm thy}$}
\startdata
$D^{2}_{A}E(z)Y,T$ & $-4.32^{+0.07}_{-0.06}$ & $2.48^{+0.20}_{-0.22}$ & $2.50$ \\           
$D^{2}_{A}E(z)^{-2/3}Y,M_{\rm t}$ & $-4.80^{+0.21}_{-0.21}$ & $1.28^{+0.27}_{-0.23}$ & $1.67$ \\
$D^{2}_{A}E(z)^{-2/3}Y,M_{\rm g}$ & $-4.89^{+0.22}_{-0.22}$ & $1.29^{+0.28}_{-0.25}$ & $1.67$ \\ 
$E(z)M_{\rm t},T     $ & $0.66^{+0.11}_{-0.12}$ & $0.95^{+0.66}_{-0.60}$ & $1.50$ \\
\enddata
\tablecomments{All cluster properties used in the analysis are based on the AMiBA SZE data (see Sec.~\ref{sec:property}),
except for the $T$ in the $M$-$T$ relation, where the $T$ is from \citet{Bonamente2007} for A1689, A1995, A2163, A2261,
and from \citet{Morandi2007} for A2390. 
The units of $T$, $D^{2}_{A}Y$, $M_{\rm t}$, and $M_{\rm g}$ are 
$7keV$, $Mpc^{2}$, $10^{14}M_{\odot}$, and $10^{13}M_{\odot}$, respectively. 
The last column $B_{\rm thy}$ indicates the theoretical values predicted by self-similar model.
In the first column, 
$E^{2}(z)\equiv\Omega_{\rm M}\left(1+z\right)^{3}+\left(1-\Omega_{\rm M}-\Omega_{\rm \Lambda}\right)\left(1+z\right)^{2}+\Omega_{\rm \Lambda}$.}
\end{deluxetable}

\section{Discussions and Conclusion}\label{sec:discuss}

We derived the cluster properties, including $T_{\rm e}$, $r_{2500}$ ,$M_{\rm t}$, $M_{\rm g}$ and $Y$,
for six massive galaxy clusters ($M_{\rm t}(r_{2500})>2\times 10^{14}M_{\odot}$) mainly based on the AMiBA SZE data. 
These results are in good agreement with those obtained solely from the OVRO/BIMA SZE data,
and those from the joint SZE-X-ray analysis of Chandra-OVRO/BIMA data.
In the comparison, the SZE-X-ray joint analysis gives smaller error bars than the pure SZE results,
because currently the uncertainty in the measurement of the SZE flux is still large.
On the other hand, in our current SZE-based analysis,
due to the insufficient $u$-$v$ coverage of the 7-element AMiBA
we still need to use X-ray parameters for the cluster model
i.e., the $\beta$ and $\theta_{\rm c}$ for the $\beta$-model.
However, \citet{apex2009} have deduced $\beta$ and $\theta_{\rm c}$ from an
APEX SZE observation alone recently.
For AMiBA, the situation will be improved when 
it expands to its 13-element configuration with 1.2m antennas \citep[AMiBA13;][]{Ho2008}, 
and thus much stronger constraints on the cluster properties than current AMiBA results are expected. 
Furthermore, with about three times higher angular solution, we should be able to estimate $\beta$ and $\theta_{\rm c}$ from our SZE data with AMiBA13
and make our analysis purely SZE based \citep{Ho2008,Sandor2008}.
Nevertheless,
the techniques of using SZE data solely to estimate cluster properties are still important,
because
many upcoming SZE surveys will observe SZE clusters for which no X-ray data are available 
\citep{Ruhl2004,Fowler2004,Kaneko2006,Ho2008},
especially for those at high redshifts.

\citet{Hallman2007} suggested that adopting the UTP $\beta$ model for SZE data on galaxy clusters
will reduce the overestimation of the integrated Compton $Y_{500}$ and gas mass. However, the $Y_{2500}$
values we obtained with the UTP model are not smaller than those obtained with the isothermal model. The $M_{g}(r_{2500})$
values deduced using the UTP model are even larger than those deduced using the isothermal model.

For the case of integrated Compton Y, when we compare $Y_{500}$ deduced using the UTP model $Y_{500,\rm UTP}$, 
and those deduced using the isothermal model $Y_{500,\rm iso}$, 
we found that the $Y_{500,\rm UTP}$ are smaller than $Y_{500,\rm iso}$, as predicted by \citet{Hallman2007}.
The reason is that the Compton $y$ profile predicted using the UTP $\beta$ model will decrease more quickly than
the profile predicted by the isothermal $\beta$ model, with increasing radius.
Therefore, the ratio $Y_{\Delta,\rm UTP}/Y_{\Delta,\rm iso}$ will decrease as $\Delta$ decreases.

We also noticed that the electron temperature values obtained with the isothermal model are significantly higher 
than the temperatures deduced from X-ray data for most clusters we considered.
The temperatures of clusters obtained using the UTP model are lower than those obtained with the isothermal model
and thus are in better agreement with those deduced from X-ray data.
Therefore, in the UTP model, with similar $Y_{2500}$ and lower temperature, we should get larger $M_{g}$.

The electron temperatures derived using the UTP $\beta$ model are in better agreement 
with X-ray observation results than those derived using the isothermal $\beta$ model.
This result implies that the UTP $\beta$ model may provide better estimates of the electron temperature when we can 
use only the $\beta$ model parameters from X-ray observation. However, we noticed that
the UTP $\beta$ model produced larger errorbars than the isothermal $\beta$ model did.
These increased errors are based on the uncertainties of $\beta$ and $r_{c}$ which we insert by hand.
On the other hand, because we treat $\beta$ and $r_{c}$ as independent parameters in this work,
the uncertainty could be over estimated due to the degeneracy between these two parameters.
If we can access to the likelihood distributions of $\beta$ and $r_{c}$ of the UTP $\beta$ model derived from observation,
the error-bars might be reduced significantly. 
 
There is a concern that 
the scaling relations among the purely SZE-derived cluster properties may be implicitly embedded 
in the formalism we used here. In this paper,
we also investigate for the first time the embedded scaling relations
between the SZE-derived cluster properties.
Our analytical and numerical analyses both suggest that there are embedded scaling relations
between SZE-derived cluster properties, with both the isothermal model and the UTP model, while we fix $\beta$. 
The embedded $Y$-$T$ and $M$-$T$ scaling relations are close to the predictions of self-similar model.
The results imply that the assumptions built in the pure-SZE method
significantly affect the scaling relation between the SZE-derived properties.
Therefore, we should treat those scaling relations carefully.

Our results suggest the possibility of measuring cluster parameters with SZE observation alone.
The agreement between our results and those from the literature provides 
not only confidence for our project
but also supports our understanding of galaxy clusters.
The upcoming expanded AMiBA with higher sensitivity and better resolution will 
significantly improve the constraints on these cluster properties.
In addition, an improved determination of the $u$-$v$ space structure of the clusters directly from AMiBA
will make it possible to measure the properties of clusters which currently do not have good X-ray data. 
The ability to estimate
cluster properties based on SZE data will improve the study of mass distribution at high redshifts.
On the other hand, the fact that the assumptions of cluster mass and temperature profiles significantly bias the estimations of scaling relations should be also noticed and treated carefully.

\acknowledgments
We thank the Ministry of Education, the National Science
Council (NSC), and the Academia Sinica, Taiwan, for their funding and supporting of AMiBA project. YWL thank the AMiBA team
for their guiding, supporting, hard working, and helpful discussions. 
We are grateful for computing support from the National Center for High-Performance Computing, Taiwan.
This work is also supported by National Center for Theoretical Science, and
Center for Theoretical Sciences, National Taiwan University for J.H.P.~Wu.
Support from the STFC for M. Birkinshaw is also acknowledged.


\end{document}